# Low temperature vortex phase diagram of $Bi_2Sr_2CaCu_2O_8$: a magnetic penetration depth study.

R. Prozorov[a], R. W. Giannetta[a], T. Tamegai[b], P. Guptasarma[c], and D.G. Hinks[c]

[a] Loomis Laboratory of Physics, University of Illinois at Urbana-Champaign,
1110 West Green St, Urbana, Illinois 61801.

[b] Department of Applied Physics, The University of Tokyo,
Hongo, Bunkyo-ku, Tokyo, 113-8656, Japan.

[c] Chemistry and Materials Science Division, Argonne National Laboratory,
Argonne, Illinois 60439

We report measurements of the magnetic penetration depth $\lambda_m(T)$ in the presence of a DC magnetic field in optimally doped *BSCCO-2212* single crystals. Warming, after magnetic field is applied to a zero-field cooled sample, results in a non-monotonic $\lambda_m(T)$, which does not coincide with a curve obtained upon field cooling, thus exhibiting a hysteretic behaviour. We discuss the possible relation of our results to the vortex decoupling, unbinding, and dimensional crossover.

## 1. INTRODUCTION

The field-temperature phase diagram of Bi-2212 is well studied at high and intermediate temperatures [1-5]. At low temperatures, the situation is less clear. Below $t=T/T_c \sim 0.2$ the fishtail disappears, persistent current density increases almost exponentially (Fig.1), and the relaxation rate changes [2]. Theory predicts various peculiarities in vortex behavior at low temperatures, such as dimensional crossover in the pinning mechanism [2], topological transition in the vortex lattice [3,4], electromagnetic decoupling and a related Kosterlitz-Thouless type transition [5].

We present new experimental results on $\lambda_m(T)$ at low temperatures and discuss their relevance to the aforementioned scenarios.

## 2. EXPERIMENTAL

Magnetic penetration depth measurements were performed using a tunnel-diode driven 11 MHz LC resonator [6] operating in a $^3He$ refrigerator. The resonance frequency shift $\Delta f = f(T) - f(T_{min})$ is related to $\Delta\lambda_m$ via $\Delta\lambda_m = -G\Delta f$, where $G$ is the sample and apparatus dependent calibration constant [7]. Magnetization was measured using a *Quantum Design* SQUID.

## 3. RESULTS

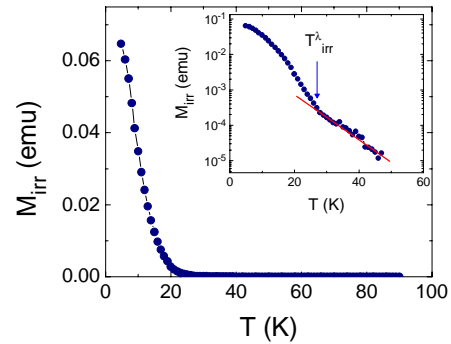

**Figure 1.** Irreversible part of the magnetic moment in *BSCCO* at $H$=300 *G*. <u>*Inset:*</u> Log plot.

Figure 1 presents the irreversible part of the magnetic moment, $M_{irr}=(M_\downarrow - M_\uparrow)/2$, as a function of temperature at $H$=300 *G*. Here, $M_\downarrow$ is the descending and $M_\uparrow$ is the ascending branch of $M(H)$ loop, respectively. The inset shows the same data in a log plot. Above certain temperature, $T_{irr}^{\lambda}$, $M_{irr}(T)$ is exponentially suppressed indicating a weak pinning regime.

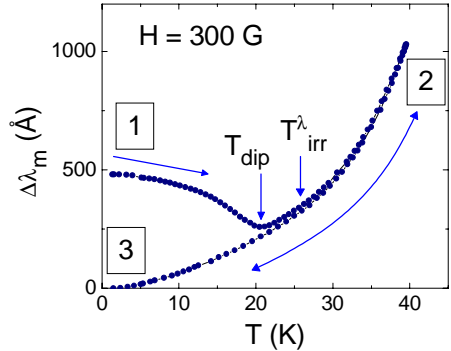

**Figure 2.** $\Delta\lambda_m(T)$ at $H$=300 $G$. Symbols and arrows are described in the text.

Figure 2 shows the magnetic penetration depth $\Delta\lambda_m(T)$ measured at $H$=300 $G$. The sample was cooled to $T$=1 $K$ and magnetic field was applied (point 1 in Fig.2). The sample was then warmed up (1→2) and cooled down (2→3). Subsequent warming and cooling did not modify the temperature dependence of $\lambda_m(T)$ – it always followed the "reversible" (3→2→3) curve. There are two distinctive points: $T_{dip}(H)$ above which $\lambda_m(T)$ is dominated by the reversible (3→2) curve; and $T_{irr}^{\lambda}(H)$ where reversible and irreversible curves merge. The observed hysteresis in $\lambda_m(T)$ can be attributed to a crossover from a strong to a weak pinning regime, which is consistent with the measurements of the irreversible magnetization in Fig. 1.

We measured $\lambda_m(T)$ at different values of the DC magnetic field and determined both $T_{dip}(H)$ and $T_{irr}^{\lambda}(H)$. The resulting phase diagram is shown in Fig.3. The usual irreversibility temperature, $T_{irr}(H)$, determined from the AC susceptibility measurements is also shown for comparison.

Unlike $T_{irr}(H)$, neither $T_{dip}(H)$ nor $T_{irr}^{\lambda}(H)$ extrapolate to $T_c$, but at most to $t$=0.5. This fact favors an unbinding transition scenario, in which the Kosterlitz - Thouless temperature sets the temperature scale [5].

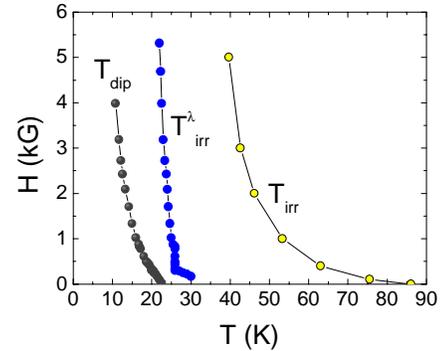

**Figure 3.** Low temperature phase diagram of BSCCO-2212 from the measurements of $\Delta\lambda_m$.

Alternative mechanisms could be a dimensional crossover in the pinning mechanism [2] or a topological transition in the vortex structure [3,4]. We also note that the fishtail feature exists only between $T_{dip}$ and $T_{irr}$ lines, Fig.3, where the pinning is weak. This would imply a collective creep, dynamic explanation of the fishtail. However, a knee in $T_{irr}^{\lambda}(H)$ at $H$≈400 $G$ (onset of a fishtail) could be an indication of the entanglement crossover [4] in the vortex structure in this temperature region.

We thank Vadim Geshkenbein for useful discussions. This work was supported by Science and Technology Center for Superconductivity Grant No. NSF-DMR 91-20000.